\documentclass{appolb}

\usepackage{graphicx}
\usepackage{amsmath,amssymb}
\usepackage{amsfonts}
\usepackage{physics}
\usepackage{xspace}
\usepackage{bm}
\usepackage{mathrsfs}

\usepackage[dvipsnames]{xcolor}
\usepackage{multirow}
\usepackage[format=plain,justification=RaggedRight,labelsep=endash,font=small,labelfont=sc]{caption}
\newcommand{\pdagger}{{\phantom{\dagger}}}

\usepackage[T1]{fontenc}

\usepackage{graphicx}
\usepackage{mathptmx}      
\usepackage{flushend}
\usepackage[numbers,sort&compress]{natbib}
\usepackage[colorlinks,citecolor=blue,urlcolor=blue,linkcolor=blue]{hyperref}

\begin{document}

\title{Quantum fluctuations of energy in subsystems of a hot relativistic gas%
}
\author{Arpan Das
\address{Institute of Nuclear Physics Polish Academy of Sciences, PL-31-342 Krakow, Poland}\\
Wojciech Florkowski
\address{Institute of Theoretical Physics, Jagiellonian University, PL-30-348 Krakow, Poland}\\
Radoslaw Ryblewski
\address{Institute of Nuclear Physics Polish Academy of Sciences, PL-31-342 Krakow, Poland}\\
Rajeev Singh
\address{Institute of Nuclear Physics Polish Academy of Sciences, PL-31-342 Krakow, Poland}}






\date{Received: date / Accepted: date}

\maketitle

\begin{abstract}
We derive a formula that defines quantum fluctuations of energy in subsystems of a hot relativistic gas. For small subsystem sizes we find substantial increase of fluctuations compared to those known from standard thermodynamic considerations. However, if the size of the subsystem is sufficiently large, we reproduce the result for energy fluctuations in the canonical ensemble. Our results are subsequently used in the context of relativistic heavy-ion collisions to introduce limitations of the concepts such as classical energy density or fluid element. In the straightforward way, our formula can be applied in other fields of physics, wherever one deals with hot and relativistic matter.
\end{abstract}

\section{Introduction}
\smallskip
Fluctuations of various physical quantities play a very important role in all fields of physics, as they reveal the information about possible phase transitions~\cite{Smoluchowski,PhysRevLett.85.2076},
formation of structures in the Early Universe~\cite{Lifshitz:1963ps,PhysRevLett.49.1110}, and dissipative phenomena~\cite{Kubo1}. Most common fluctuations we deal with are those arising from quantum uncertainty relation or those present in thermodynamic systems~\cite{Huang:1987asp}. Our present study combines these two physics aspects --- we study quantum fluctuations of energy in subsystems of a hot relativistic gas and demonstrate that they agree with thermodynamic fluctuations in canonical ensemble if subsystems are sufficiently large. On the other hand, we find a very substantial increase of the fluctuations for small subsystems. 

The results of our numerical calculations may serve as a guide that shows, for a given temperature and particle mass, how large the size of the subsystem is for which the quantum fluctuations of energy become classical and eventually may be neglected. This, in turn, may be used to validate the concept of energy density used in classical description of fluids.

The physical system we are particularly interested in is hot matter produced in relativistic heavy-ion collisions. The concept of a hot gas is dominantly used here in various aspects: from perturbative quark-gluon-plasma possibly produced at the top beam energies to hadron resonance gas produced at hadronic freeze-out. Combined with the concept of hot gas is the use of relativistic hydrodynamics which uses equation of state that is very often of the ideal-gas form. 

In fact, relativistic hydrodynamics has become now the main theoretical tool used to interpret heavy-ion collisions. Its overwhelming successes paved the road to its wider and wider applications --- in the regions where its applicability range might be questioned~\cite{Jaiswal:2016hex,Florkowski:2017olj,Romatschke:2017ejr}. The basic concepts of energy density and pressure are used now within hydrodynamics to characterize very small portions of matter at extreme conditions. This leads us back to the question of how well the energy of such small systems is well defined and how seriously we can take hydrodynamic structures of the energy density profiles which vary on a space scale that is a fraction of one Fermi.

The results presented below have two aspects: We first derive a compact formula that defines quantum fluctuations in a subsystem of a hot relativistic gas, then we apply this formula in the physical situations expected in relativistic heavy-ion collisions. Below we use the West--Coast metric $g_{\mu\nu} = \hbox{diag}(+1,-1,-1,-1)$. Three-vectors are shown in bold font and a dot is used to denote the scalar product of both four- and three-vectors, i.e., $a^\mu b_\mu = a \cdot b = a^0 b^0 - \boldsymbol{a} \cdot  \boldsymbol{b}$.
\smallskip
\section{Basic concepts and definitions}
\smallskip
In this work we consider a subsystem $S_a$ of the thermodynamic system $S_V$ described by the canonical ensemble characterized by the temperature $T$ (or its inverse $\beta = 1/T$). For sake of simplicity, we assume that the system $S_V$ consists of spinless boson particles with mass $m$. The characteristic volume of the system $S_a$ is always smaller than the volume $V$ of the system $S_V$, and $V$ is sufficiently large to allow for doing integrals over particle momenta (instead of sums imposed by otherwise commonly used box periodic conditions). 

With these assumptions in mind, we describe our system by a quantum scalar field in thermal equilibrium. The field operator has the standard form~\cite{Chen:2018cts}
\begin{equation}
\phi(t,{\boldsymbol{x}})=\int\frac{d^3k}{\sqrt{(2\pi)^3 ~2\omega_{\boldsymbol{k}}}}\left(a_{\boldsymbol{k}}^{\pdagger}e^{-i k \cdot x} +
a_{\boldsymbol{k}}^{\dagger}e^{i k \cdot x} \right),
\label{equ1ver1}
\end{equation}
where $a_{\boldsymbol{k}}^{\pdagger}$ and $a_{\boldsymbol{k}}^{\dagger}$ are annihilation and creation operators, respectively, satisfying the canonical commutation relations $[a_{\boldsymbol{k}}^{\pdagger},a_{\boldsymbol{k}^{\prime}}^{\dagger}] = \delta^{(3)}(\boldsymbol{k}-\boldsymbol{k}^{\prime})$, whereas $\omega_{\boldsymbol{k}}=\sqrt{{\boldsymbol{k}}^2+m^2}$ is the energy of a particle. To perform thermal averaging, it is sufficient to know the expectation values of the products of two and four creation and/or annihilation operators~\cite{CohenTannoudji:422962,Itzykson:1980rh,Evans:1996bha}
%
\begin{align}
\langle a^{\dagger}_{{\boldsymbol{k}}}a_{{\boldsymbol{k}}^{\prime}}^{\pdagger}\rangle&=\delta^{(3)}({\boldsymbol{k}}-{\boldsymbol{k}}^{\prime})f(\omega_{{\boldsymbol{k}}}),\label{equ2ver1}\\
\langle a^{\dagger}_{{\boldsymbol{k}}}a^{\dagger}_{{\boldsymbol{k}}^{\prime}}a_{{\boldsymbol{p}}}^{\pdagger}a_{{\boldsymbol{p}}^{\prime}}^{\pdagger}\rangle &= \bigg(\delta^{(3)}({\boldsymbol{k}}-{\boldsymbol{p}})~\delta^{(3)}({\boldsymbol{k}}^{\prime}-{\boldsymbol{p}}^{\prime}) \label{equ3ver1} \\
&+\delta^{(3)}({\boldsymbol{k}}-{\boldsymbol{p^{\prime}}})~\delta^{(3)}({\boldsymbol{k}}^{\prime}-{\boldsymbol{p}})\bigg)f(\omega_{{\boldsymbol{k}}})f(\omega_{{\boldsymbol{k}}^{\prime}}). \nonumber 
\end{align}
Here $f(\omega_{{\boldsymbol{k}}})$ is the Bose--Einstein distribution function, $f(\omega_{{\boldsymbol{k}}})=1/(\exp[\beta ~\omega_{{\boldsymbol{k}}}]-1)$. Any other combinations of two and four creation and/or annihilation operators can be obtained from Eqs.~\eqref{equ2ver1} and \eqref{equ3ver1}
through the commutation relation between $a_{{\boldsymbol{k}}}^{\pdagger}$ and $a_{{\boldsymbol{k}}}^{\dagger}$.

Following~\cite{Chen:2018cts}, we define an operator $\mathcal{H}_a$ that describes the energy density of a {\it finite} subsystem $S_a$ placed at the origin of the coordinate system 
\begin{align}
\mathcal{H}_a = \frac{1}{(a\sqrt{\pi})^3}\int d^3{\boldsymbol{x}}~\mathcal{H}(x)~\exp\left(-\frac{{\boldsymbol{x}}^2}{a^2}\right).
\label{equ4ver1}
\end{align}
Here $\mathcal{H}(x) \equiv (\dot{\phi}^2+({\bf{\nabla}\phi)}^2+m^2\phi^2)/2$ is the Hamiltonian density of the free real scalar field. The Gaussian profile used in Eq.~\eqref{equ4ver1} is in fact our definition of the subsystem $S_a$ -- the smooth profile with a length scale $a$ has been introduced instead of a cube to remove possible boundary effects coming from sharp boundaries. 

The thermal expectation value of the operator $ \mathcal{H}_a$ is 
\begin{align}
    \langle :\mathcal{H}_a :\rangle= \int \frac{d^3{{k}}}{(2\pi)^3}~\omega_{{\boldsymbol{k}}}~f\left(\omega_{{\boldsymbol{k}}}\right) \equiv \varepsilon(T).
    \label{equ5ver1}
\end{align}
This is certainly an expected result known from elementary kinetic-theory considerations~\cite{Huang:1987asp}. To remove an infinite vacuum part coming from zero-point energy contributions, we have applied standard normal ordering procedure to $\mathcal{H}_a$ on the left-hand side of Eq.~\eqref{equ5ver1}. We note that the energy density $\varepsilon$ defined by Eq.~\eqref{equ5ver1} is independent of $a$, which reflects the spatial uniformity of the system $S_V$. Our results are also independent of time. 

To determine the fluctuation of energy of the subsystem $S_a$, in agreement with general quantum-mechanics rules we consider the variation
\begin{equation}
 \sigma^2(a,m,T) = \langle :\mathcal{H}_a: :\mathcal{H}_a: \rangle - \langle :\mathcal{H}_a :\rangle^2\, 
 \label{sigma2}
\end{equation}
or the normalized standard deviation
\begin{equation} 
\sigma_n(a,m,T)= \frac{(\langle:\mathcal{H}_a::\mathcal{H}_a:\rangle- \langle :\mathcal{H}_a :\rangle^2)^{1/2}}{\langle :\mathcal{H}_a :\rangle}.
\end{equation}

Using the thermal expectation values of the products of two and four creation and annihilation operators, as given by Eqs.~\eqref{equ2ver1} and \eqref{equ3ver1},  we find
\begin{align}
\sigma^2(a,m,T) &=  \int dK ~dK^{\prime} f(\omega_{{\boldsymbol{k}}})(1+f(\omega_{{\boldsymbol{k}}^{\prime}}))\nonumber\\
&\times \bigg[(\omega_{{\boldsymbol{k}}}\omega_{{\boldsymbol{k}}^{\prime}}+{\boldsymbol{k}}\cdot{\boldsymbol{k}}^{\prime}+m^2)^2e^{-\frac{a^2}{2}({\boldsymbol{k}}-{\boldsymbol{k}}^{\prime})^2}\nonumber\\
&+(\omega_{{\boldsymbol{k}}}\omega_{{\boldsymbol{k}}^{\prime}}+{\boldsymbol{k}}\cdot{\boldsymbol{k}}^{\prime}-m^2)^2e^{-\frac{a^2}{2}({\boldsymbol{k}}+{\boldsymbol{k}}^{\prime})^2}\bigg].
\label{equ6ver1}
\end{align}
where $dK = d^3{{k}}/((2\pi)^{3} 2 \omega_{{\boldsymbol{k}}})$. In Eq.~\eqref{equ6ver1} we have discarded a divergent term that is temperature independent and may be attributed to the pure vacuum energy fluctuation. As a matter of fact, the latter was studied in Ref.~\cite{Phillips:2000jm}, where a technique of smeared and displaced operators was used. The variation of the energy density divided by the mean energy density squared was found in this case to be 2/3. As this result corresponds to the limit $a \to 0$ for $m=0$, it is negligible compared to our value of $\sigma^2_n$ that diverges as $a \to 0$. On the other hand, using dimensional analysis one expects that the variation of the vacuum energy density for massless particles in a volume of the size $a^3$ should decrease as $1/a^8$, which makes it again smaller compared to our effects that decrease as $T^5/a^3$, which will be demonstrated below. Altogether, we expect that the inclusion of the properly regularized vacuum fluctuations into Eq.~\eqref{equ6ver1} cannot quantitatively alter our conclusions.


Equation~\eqref{equ6ver1} represents our main result that allows us to determine the energy fluctuations of ``Gaussian'' subsystems $S_a$ of the system $S_V$. By numerical integration we may obtain the results for any subsystem size $a$, temperature $T$, and for particle mass $m$. Nevertheless, before we present our numerical results, it is instructive to consider the limiting case of very large system size $a$ and to analyze special cases for which analytic results can be obtained. It is also useful to generalize our formulas to the case where we deal with several particle types.

\smallskip
\section{Degeneracy factor}
\smallskip
Studying thermodynamic properties of particles one usually introduces degeneracy factors connected with so-called internal degrees of freedom such as spin, isospin or color charge. In our approach, we have so far considered only one type of boson particle without any internal quantum numbers. To take into account $g$ copies of such particles,  one has to include $g$ copies of the scalar field expressed by the creation and annihilation operators that commute for different particle species. Altogether this procedure results in the simple replacements
\begin{equation}
\varepsilon \rightarrow g \varepsilon, \quad
\sigma^2 \rightarrow g \sigma^2.
\label{g}
\end{equation}
As the multiplication of the energy density by $g$ is rather straightforward, the scaling of $\sigma^2$ with $g$ is less obvious --- it is a consequence of the fact that the mixed terms describing different types of particles in Eq.~\eqref{sigma2} vanish.
\begin{figure}[t]
\begin{center}
	\includegraphics[scale=0.45]{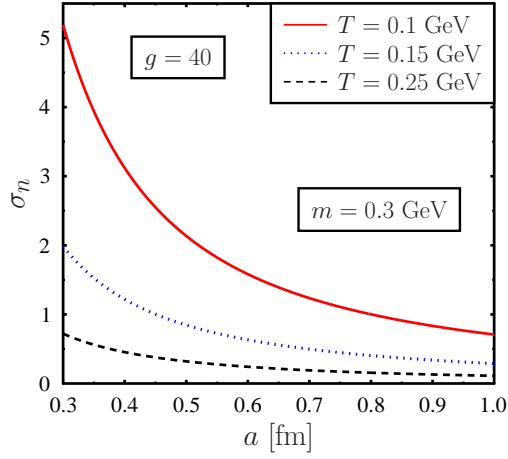}
	\caption{Variation of the normalized energy density fluctuation $\sigma_n$ in the subsystem $S_a$ with the length scale $a$ for different values of the temperature $T$ and fixed particle mass $m=0.3$ GeV.}
	\label{fig:m_0.3}
\end{center}
\end{figure}
\begin{figure}[t]
\begin{center}
	\includegraphics[scale=0.45]{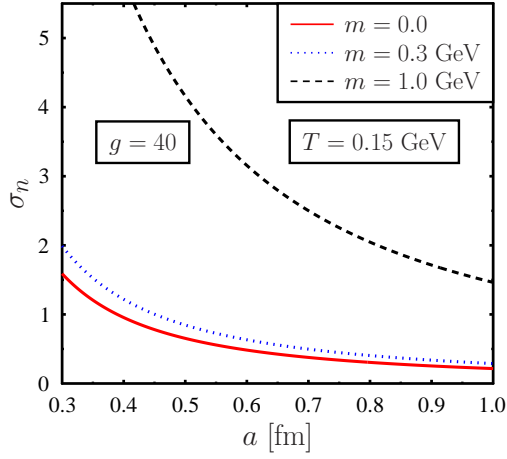}
	\caption{Same as Fig. 1 but for fixed temperature $T~=~0.15$~GeV and different particles masses.}
	\label{fig:T_0.15}
	\end{center}
\end{figure}
\section{Thermodynamic limit}
As we have mentioned earlier, $S_a$ is a subsystem of the system $S_V$. We expect that in the limit 
$a\rightarrow \infty$ (still with $a^3 \ll V$) our formula for quantum fluctuation is reduced to that known from classical statistical mechanics~\cite{Huang:1987asp}. To check this property we use in Eq.~\eqref{equ6ver1} the Gaussian representation of the three dimensional Dirac delta function
\begin{align}
    \delta^{(3)}({\boldsymbol{k}}-{\boldsymbol{p}})=\lim_{a \to\infty} \frac{a^3}{(2\pi)^{3/2}}e^{-\frac{a^2}{2}({\boldsymbol{k}}-{\boldsymbol{p}})^2}.
\end{align}
This leads us to the formula valid in the large $a$ limit
\begin{align}
\sigma^2 & \sim 
\frac{g}{(2\pi)^{3/2} a^3}
\int \frac{d^3{{k}}}{(2\pi)^3}~\omega_{{\boldsymbol{k}}}^2~f(\omega_{{\boldsymbol{k}}}) (1+f(\omega_{{\boldsymbol{k}}})).
\nonumber
\end{align}
We note that for massless particles $\sigma^2 \sim T^5/a^3$, which is the result mentioned above. The right-hand side of the last equation can be expressed in terms of the specific heat at  constant volume
\begin{equation} 
c_V = \frac{d\varepsilon}{dT} = \frac{g}{T^2} \int \frac{d^3{{k}}}{(2\pi)^3}~\omega_{{\boldsymbol{k}}}^2~f(\omega_{{\boldsymbol{k}}}) (1+f(\omega_{{\boldsymbol{k}}})).
\end{equation}
Therefore, we find (again in the large $a$ limit) 
\begin{align}
V_a \sigma_n^2
= \frac{T^2 c_V}{\varepsilon^2}
= V \frac{\langle H^2\rangle-\langle H \rangle^2}{\langle H \rangle^2} \equiv V \sigma^2_H,
\label{equ10ver1}
\end{align}
where $V_a = a^3 (2\pi)^{3/2}$ and $H$ is the Hamiltonian of $S_V$. The right-hand side of Eq.~\eqref{equ10ver1} may be identified as the normalized energy fluctuation in the system $S_V$~\cite{Huang:1987asp}. One may identify $V_a=a^3 (2\pi)^{3/2}$ as the volume of the subsystem $S_a$ --- a nontrivial factor of $(2\pi)^{3/2}$ is an artifact of using the ``Gaussian'' box. We note that Eq.~\eqref{equ10ver1} is also consistent with the result obtained in Ref.~\cite{Mrowczynski:1997mj}, where purely classical concepts have been used. 
%
 \section{Massless case with Boltzmann statistics}
In principle, the last result can be alternatively shown by first finding an analytic expression for  $V_a \sigma_n^2$ and then finding its asymptotic form for $a\rightarrow \infty$ limit. Unfortunately, in the general case of finite particle mass and Bose-Einstein statistics, the integral \eqref{equ6ver1} is not analytic and such procedure cannot be easily accomplished. However, the integrals appearing in Eq. \eqref{equ6ver1} become analytic for the Boltzmann statistics and $m=0$. In this case we obtain
\begin{equation}
\langle :\mathcal{H}_a :\rangle^2 = \frac{9 g^2}{\pi^4} T^8\, ,
\end{equation} 
and
\begin{align}
\sigma_n^2 &= \frac{1}{4320 \,g}\bigg[2970 ~\zeta^4-540~\zeta^6-96 ~\zeta^8-28~\zeta^{10}\nonumber\\
&-2 ~\zeta^{12}+\sqrt{2\pi}~e^{\frac{\zeta^2}{2}}\bigg(1485~\zeta^3-765~\zeta^5+ 300~\zeta^7\nonumber\\
&+60~\zeta^9+15~\zeta^{11}+ \zeta^{13}\bigg)\text{erfc}\bigg(\frac{\zeta}{\sqrt{2}}\bigg)\bigg],
\label{equ14ver1}
\end{align}
where $\zeta = 1/(aT)$. Using this result we find  
\begin{align}
\lim_{a \to\infty} V_a \sigma_n^2 =  \frac{11 \pi^2}{8g}\frac{1}{T^3},
\end{align}
which is the normalized energy fluctuation $V \sigma^2_H$ of the system $S_V$ as can be obtained from Eq.~\eqref{equ10ver1} for Boltzmann distribution functions with $m=0$. We emphasize that for $a\rightarrow\infty$ the analytic result for the case $m=0$ and Boltzmann statistics agrees with the result obtained earlier with the Gaussian approximation for the Dirac delta function.
\section{Numerical results} 
Having checked that our expression for quantum fluctuation correctly reproduces the thermodynamic limit for $a\rightarrow\infty$ and with the massless-Boltzmann case known analytically, we can present the results of our numerical calculations. 
In Figs.~\eqref{fig:m_0.3} and \eqref{fig:T_0.15} we show the variation of the normalized energy density fluctuation $\sigma_n$ with the subsystem size $a$ for different values of temperature and mass, respectively. With possible interpretation of heavy-ion data in mind, we consider temperatures in the range 100~MeV $< T <$ 250~MeV, and particle masses: $m~=~0,~300$ and 1000~MeV. The effective number of degrees of freedom in a perturbative quark-gluon-plasma varies between 37 (for two quark flavors) and 47.5 (for three quark flavors). In a hot hadron gas used to describe the properties of matter at freeze-out one includes about 400 hadronic states ~\cite{Kisiel:2005hn} but these are usually very heavy and equivalent to a smaller number of lighter particles (in terms of energy density or pressure). In order to get some rough estimate of $g$ we use the value $g=40$.

Figures~\eqref{fig:m_0.3} and \eqref{fig:T_0.15} show that $\sigma_n$ decreases with increasing size of the system $a$, which is the expected general behavior of fluctuations. By the way, we note that Eq.~\eqref{equ6ver1} implies that $\sigma_n$ diverges in the limit $a \to 0$ (which is a consequence of uncertainty relation), so our plots start only at $a=0.3$~fm. For fixed particle mass, the normalized fluctuations decrease with growing temperature. On the other hand, at fixed $T$ the fluctuations grow with $m$. Interestingly, the standard deviations alone exhibit an opposite behavior (not shown here).

\begin{figure}[t]
\begin{center}
\includegraphics[scale=0.45]{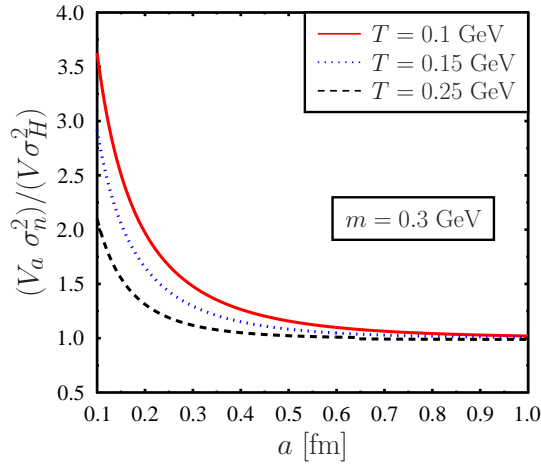}
\caption{Variation of the normalized energy fluctuation in the subsystem $S_a$ with the length scale $a$ for different values of the temperature at $m=0.3$ GeV.
}
\label{fig:norm_0.3}
\end{center}
\end{figure}
\begin{figure}[t]
\begin{center}
	\includegraphics[scale=0.45]{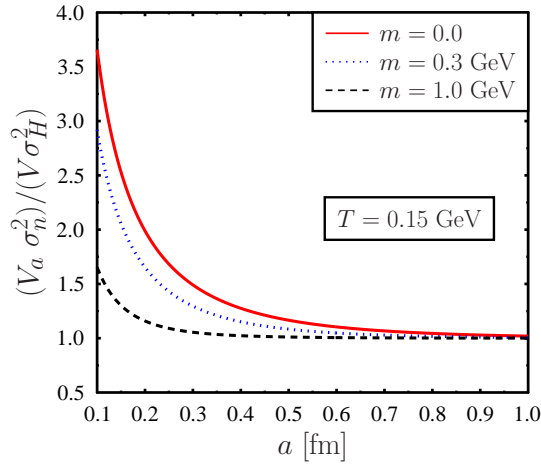}
	\caption{Variation of the normalized energy fluctuation in the subsystem $S_a$ with the length scale $a$ for different values of mass at $T=0.15$ GeV.
	}
	\label{fig:norT_0.15}
	\end{center}
\end{figure}

We have analytically demonstrated above that the volume scaled normalized fluctuation $V_a \sigma_n^2$ approaches thermodynamic limit in the large volume limit for massless particles. Figs.~\eqref{fig:norm_0.3} and \eqref{fig:norT_0.15} present the variation of $V_a \sigma_n^2/V\sigma^2_H$ with the size of the subsystem $S_a$ in the case where particles have a non vanishing mass and they obey Bose-Einstein statistics. The values of the parameters are the same as in Figs.~\eqref{fig:m_0.3} and \eqref{fig:T_0.15}. From Eq.~\eqref{equ10ver1} one expects that in the thermodynamic limit $V_a\sigma_n^2/V\sigma^2_H$ should approach unity. This property is nicely seen in Figs.~\eqref{fig:norm_0.3} and \eqref{fig:norT_0.15}, where we observe that the quantum fluctuations agree with the thermodynamic ones already for $a >$~1 fm. On the other hand, the quantum fluctuations become very important at the scale of 0.1 fm, which excludes classical treatment of such small subsystems. It also suggests that quantum fluctuation should be combined in future works with hydrodynamic fluctuations~\cite{Kapusta:2011gt}. 

For given values of $T$, $m$, and $g$, our results may be used to quickly check if the corresponding fluctuations (quantum and thermodynamic ones) in a finite subsystem of the size $a^3$ are sufficiently small, let us say, smaller than 1. If this condition is not satisfied, the classical picture of a well defined energy density in the gas/fluid cells of the size $a^3$ is not well defined. For example, for massless particles at the temperature 150 MeV (with $g=40$), the normalized fluctuations become larger than unity for $a < 0.4$~fm. This suggests that for smaller sizes the effects of fluctuations become relevant for the system's description. The classical description with ``well defined energy density'' makes sense only after coarse graining over the scales larger that 0.4 fm. 

\section{Conclusions} 
In this work we have derived the formula characterizing the quantum fluctuation of energy in subsystems of a hot relativistic gas. We have shown that it agrees with the expression for thermodynamic fluctuations, if the size of the subsystem is sufficiently large. We have found exact analytic expression for the energy fluctuation in the case of massless particles described by Boltzmann statistics. The consequences one can draw from our formula for  description of hot and relativistic systems have been discussed in the case of heavy-ion physics. Applications of our approach to other systems are straightforward.\\

We would like to thank Krzysztof Golec-Biernat and Mariusz Sadzikowski for very useful and illuminating discussions. This research was supported in part by the Polish National Science Centre Grants No. 2016/23/B/ST2/00717 and No. 2018/30/E/ST2/00432.

\bibliography{fluctuationRef.bib}{}
\bibliographystyle{spphys}

\end{document}